\documentstyle[mymulticol,pra,aps,amsfonts]{revtex}

\newcommand{\cR}{{\cal R}}

\newcommand{\beq}{\begin{equation}}
\newcommand{\eeq}{\end{equation}}
\newcommand{\beqy}{\begin{eqnarray}}
\newcommand{\eeqy}{\end{eqnarray}}

\newenvironment{Definition*}{{\bf Definition}}{}
 
\makeatletter
\def\@beginTheorem#1#2{\trivlist \item[\hskip \labelsep{\bf #1\ #2}]}
\def\@opargbegintheorem#1#2#3{ \trivlist
      \item[\hskip \labelsep{\bf #1\ #2\ (#3)}]}
\makeatother

\makeatletter
\def\@beginLemma#1#2{\trivlist \item[\hskip \labelsep{\bf #1\ #2}]}
\def\@opargbeginLemma#1#2#3{ \trivlist
      \item[\hski
 Hence we have the same statements
about the increase of the supports for increasing depth, where
the local transformations are not counted for the depth.
p \labelsep{\bf #1\ #2\ (#3)}]}
\makeatother

\makeatletter
\def\@beginDefinition#1#2{\trivlist \item[\hskip \labelsep{\bf #1\ #2}]}
\def\@opargbeginDefinition#1#2#3{ \trivlist
      \item[\hskip \labelsep{\bf #1\ #2\ (#3)}]}
\makeatother

\makeatletter
\def\@beginCorollary#1#2{\trivlist \item[\hskip \labelsep{\bf #1\ #2}]}
\def\@opargbeginCorollary#1#2#3{ \trivlist
      \item[\hskip \labelsep{\bf #1\ #2\ (#3)}]}
\makeatother

\makeatletter
\def\@beginExample#1#2{\trivlist \item[\hskip \labelsep{\bf #1\ #2}]}
\def\@opargbeginExample#1#2#3{ \trivlist
      \item[\hskip \labelsep{\bf #1\ #2\ (#3)}]}
\makeatother

\def\R{{\mathbb{R}}}

\newcommand{\cH}{{\cal H}}

\title{Quantum algorithm for finding
periodicities in the spectrum  \newline of a
black-box  Hamiltonian or unitary transformation}

\author{D. Janzing\thanks{Electronic address: janzing@ira.uka.de}  and 
Th. Beth}
\address{Institut f\"ur Algorithmen und Kognitive Systeme, Am Fasanengarten 3a,
    D--76\,131 Karlsruhe, Germany}

\begin{document}
\maketitle

\begin{abstract}
Estimating the eigenvalues of a unitary  transformation $U$ 
by standard  phase estimation 
requires the implementation of controlled-$U$-gates which
are not available if  $U$ is  only  given as a black box.

We show that a simple trick allows to measure
eigenvalues of $U\otimes U^\dagger$ 
even in this case.
Running the algorithm several times allows therefore 
to estimate the autocorrelation function
of the density of eigenstates of $U$.
This can be applied to
find periodicities in the energy spectrum of a quantum system with
unknown Hamiltonian if it can be  coupled  to a quantum computer.
\end{abstract}

\begin{multicols}{2}

\section{Standard  phase estimation and its weakness}

Finding the eigenvalues of unitary transformations or self-adjoint operators
is a central task in quantum mechanics. The thermodynamic and dynamical
properties of a quantum system are determined by the spectrum of 
a Hamiltonian and the corresponding unitary transformations.
Furthermore the estimation of eigenvalues is an important
tool in quantum computation
(see \cite{NC}). For that reason the algorithm for {\it phase estimation}
has been developed (see \cite{NC,TMi} and references therein). We rephrase it as follows.
We have a Hilbert space $\cR_a \otimes \cH$ where 
$\cH$ is the {\it target} register where the considered unitary $U$ acts on
and an ancilla register $\cR_a$ 
consisting of $k$ qubits if an accuracy
of the eigenvalues of $U$ of the order $2^{-k}$ is desired. Assume the target 
register to be in an eigenstate  $|\psi_{\cH}\rangle$ 
of $U$ with eigenvalue
$\exp(i\phi)$. Initialize the ancilla register in an equal  superposition of
all its logical states, i.e.,
\[
|\psi_{\cR}\rangle:=  (\frac{1}{\sqrt{2}}(|0\rangle + |1\rangle))^{\otimes n}=
\frac{1}{\sqrt{2^k}}\sum_{l<2^k} |l\rangle,
\]
where $l$ is the binary number 
corresponding to the $k$ ancilla qubits $j=0,\dots,k-1$.
On the joint Hilbert space $\cR_a \otimes \cH$ 
apply for all $j=0,\dots, k-1$ the 
transformations
\begin{equation}\label{vau}
V_j:=|1_j\rangle \langle 1_j|\otimes U^{2^j} + |0_j\rangle \langle 0_j| \otimes 1,
\end{equation}
where $|1_j\rangle \langle 1_j|$ and  $|0_j\rangle \langle 0_j|$
are the projectors onto the $|0\rangle$ and $|1\rangle$ state of the ancilla 
qubit $j$,
respectively. The operation $U^{2^j}$ is the $2^j$-fold  iteration
of $U$.

If $\cH$ is in the state $|\psi_\cH\rangle$ then 
$|\psi_\cR\rangle$ is converted into the state
\[
\frac{1}{\sqrt{2^k}}\sum_l e^{i \phi l}|l\rangle
\]
by the `kick-back-effect' \cite{CEMM}.
After Fourier transformation on $\cR_a$ we obtain
\[
\frac{1}{\sqrt{2^k}}\sum_{l,m} e^{-2\pi ilm/(2^k)} e^{i \phi m}|m\rangle,
\] 
i.e., the probability distribution is peaked around $m=\phi 2^k/ \pi $.
If $|\psi_{\cH}\rangle$ is not an eigenstate of $U$, than the algorithm
will project approximatively (in the limit of large $k$) onto any
of the eigenstates \cite{TMi}. 
If the initial state on $\cH$ is a density matrix which is diagonal
in the basis of $U$, one will obtain any of the eigenvalues of $U$ with
the corresponding probability. 

At first sight, quantum phase estimation seems to be applicable
for finding energy values and eigenstates of an unknown Hamiltonian
of an arbitrary quantum system simply by setting 
 $U:=\exp(-iHt)$. This would be interesting for
the investigation of complex physical systems. 
To measure eigenvalues of 
interaction Hamiltonians in many-spin systems,
as molecules or solid states, for instance, would be rather useful.
 
But there is a severe problem which is
essentially that quantum phase estimation does {\it not} use 
the implementation
of $U$ as
a {\it black box} subroutine. 
It uses the  {\it conditional} transformations $V_j$  (see eq. (\ref{vau}))
as black boxes and one should emphasize that no canonical conversion
procedure building $V_j$ from $U$ is known if $U$ is a black box.
Sometimes this fact is hidden by using a language too classical
if one explains the action of $V_j$ 
 by claiming that it implements $U$ (or its iterations) if
the corresponding ancilla qubit is in the logical state $|1\rangle$. This
hides the fact that a superposition state of the ancilla has to lead to a 
superposition of the two actions `implementation of $U$' and 
`no implementation'. Imagine that the black box implementing $U$ contains
a memory storing the information whether $U$ is implemented or not. Then
such a superposition of implementation is destroyed by the memory.
However, in \cite{Janz} we have shown that the unitary evolution $U$
can in principle  be conjugated by other unitary transformations in such a 
way that the net effect is a `controlled-$U$'. 
But the class of unitary transformations $U$ considered there is only
the evolutions according to $n$-qubit pair-interaction Hamiltonians.
Here we address the question how to use quantum phase estimation
for obtaining information about the spectrum  of $U$ if we have no prior
information about $U$ at all. We show that it is at least possible 
to get the autocorrelation function of the spectrum of $U$, or, speaking
more explicitly, the spectrum of $U\otimes U^\dagger$, provided that the
following assumptions are true:

\begin{enumerate}
\item
The operation $U$ is implementable on a system
$\cH$ which can be brought into interaction with another
register $\cR_1$ of equal  Hilbert space dimension in such a way that
complete exchange of quantum information between $\cH$ and $\cR_1$ is possible.
If $U=\exp(-iHt)$ for an appropriate $t>0$ and $H$ is the
real Hamiltonian of the system $\cH$ we assume that this exchange of
information can either be done on a small time scale compared to the evolution
according to $H$ or the natural evolution can be switched off
during the implementation of this information exchange. 

\item
There is another quantum register $\cR_2$ with the same dimension as
$\cH$ and $\cR_1$ and an ancilla register $\cR_a$ consisting
of $k$ bits if the  desired  accuracy  for the eigenvalues is $2^k$.

\item
On the system $\cR_1\otimes \cR_2\otimes \cR_a$ we have a set
of quantum transformations available which is universal for quantum 
computation.

\end{enumerate}

Of course 
the assumption that the unknown Hamiltonian $H$ can be switched off 
is problematic, but if additional prior information about the 
structure of $H$ is available, standard decoupling techniques \cite{Le}
can be used. Note that
assumption 1 is considerably weaker than the assumption that
$H$ can be switched on and off by the quantum state of an ancilla 
qubit\footnote{This is discussed in more detail in \cite{Janz}}.

\section{Implementing a conditional transformation}

The essential part of our algorithm is rather simple in contrast 
to \cite{Janz} for the cost that we 
obtain eigenvectors and eigenvalues of $U\otimes U^{-1}$ instead of those of 
$U$. 
It consists of a conjugation of $U$ by known unitary transformations
in such a way that the net effect is the conditional transformation
\[
V'_j:= |1_j\rangle \langle 1_j|\otimes  U^{2^j}\otimes 1 + 
|0_j\rangle \langle 0_j|\otimes 
1 \otimes U^{2^j},
\]
where $|0_j\rangle$ and $|1_j\rangle$ are states of the
ancilla qubit $j$.
One can see easily that the effect on the ancilla's states
is the same if the unconditional unitary 
$U^{-2^j}$ is implemented on $\cR_2$ after each implementation
of $V_j'$. This implements the conditional transformation
\[
V_j'':= |1_j\rangle \langle 1_j|\otimes  U^{2^j}\otimes U^{-2^j} + 
|0_j\rangle \langle 0_j|\otimes 
1 \otimes 1 \,.
\]

Using standard phase estimation we can use $V_j''$ for 
obtaining eigenvalues of
$U\otimes U^\dagger$.

The  procedure for implementing $V'_j$ consists of the
following steps for  $j=0,\dots,k-1$.

\begin{enumerate}

\item
Implement state exchange of the registers $\cH$ and $\cR_1$, i.e.,
the unitary $W$ with $W (|\alpha \rangle 
\otimes |\beta\rangle):=|\beta \rangle 
\otimes |\alpha\rangle$.

\item
Implement $U^{2^j}$ on $\cH$. If $U=\exp(-iHt)$, i.e., if 
$u$ is the unitary
evolution according to the system Hamiltonian $H$, 
then one has to wait the time $2^j t$.

\item
Implement state exchange of $\cH$  and $\cR_1$ again. Steps 1-3 implement
the transformation $\exp(-iHt)$ on the register $\cR_1$.

\item
Implement a {\it conditional} exchange of the states of $\cR_1$ and $\cR_2$
depending on the state of qubit $j$ in the ancilla state, i.e. we implement
\[
|1_j\rangle \langle 1_j| \otimes  W +|0_j\rangle \langle 0_j| \otimes 1
\]
on $\cR_a\otimes \cR_1\otimes \cR_2$.

\end{enumerate}

The conditional exchange can easily be implemented, if the registers
$\cR_1$ and $\cR_2$ consist of qubits. In this case the conditioned 
permutation of two 
corresponding qubits is a usual Fredkin-gate \cite{NC}. 

Our algorithm might have applications for investigating the spectrum
of a many-particle Hamiltonian in solid-state physics, since the spectrum
and its gaps determines dynamical and thermo-dynamical behavior of the 
system \cite{AsMe}.
Of course for many-particle systems
it is not possible to find the complete set of eigenvalues
of $\exp(-iHt)\otimes \exp(iHt)$ since the dimension of $\cH$ 
grows exponentially. 
But periodicities in the spectrum of $H$ could be
found. We sketch this idea.
First we assume that $t$ is chosen in such a way that $t \Delta  \leq
\pi$, where $\Delta$ is an upper bound on the difference between
the largest and smallest eigenvalue of $H$ given by prior knowledge. 
Assume that $\cR_1$ and $\cR_2$ are prepared in the same initial
density matrix $\rho$. The state $\rho$ defines a probability measure
on $\R$ by setting $p(\lambda)=0$ if $\lambda$ is no eigenvalue of $u$
and $p(\lambda):=\langle \lambda | \rho |\lambda \rangle$ else.
If $\rho$ is the maximally mixed state and the dimension of $\cH$ is
large such that the probability measure can approximatively described by a 
probability density, this measure is known as the {\it density of states} 
\cite{Lon}. Running the algorithm several times allows to estimate the 
density of eigenstates of $H\otimes 1 -1 \otimes H$ which is the
autocorrelation function of the density of eigenstates of $H$.
If this density contains  periodicities, for instance when 
there are  spectral gaps with equal distances, they can be detected
by using our algorithm. Note that  
in solid-state physics, for instance, energy gaps occur  
which do not depend on $n$. Hence the size of the required ancilla 
register does not necessarily grow with $n$ for detecting 
interesting gaps.

\end{multicols}

\section*{Acknowledgments}

Thanks to P. Wocjan for useful remarks.
Part of this work has been supported by the European project
Q-ACTA and the DFG-project `verlustarme Informationsverarbeitung'.

\end{document}